\pdfoutput=1
\documentclass{amsart}%
\usepackage{amsfonts}
\usepackage{amsmath}
\usepackage{amssymb}
\usepackage{graphicx}%
\providecommand{\U}[1]{\protect\rule{.1in}{.1in}}

\theoremstyle{plain}

\newtheorem{definition}{Definition}
\newtheorem{example}{Example}

\newtheorem{proposition}{Proposition}
\newtheorem{remark}{Remark}

\numberwithin{equation}{section}
\begin{document}
\title[Quantizing Braids and Other Mathematical Objects]{Quantizing Braids and Other Mathematical Objects: \ The General Quantization Procedure}
\author{Samuel J. Lomonaco}
\address{University of Maryland Baltimore County (UMBC)\\
Baltimore, MD \ 21250 \ \ USA}
\email{lomonaco@umbc.edu}
\urladdr{http://www.csee.umbc.edu/\symbol{126}lomonaco}
\author{Louis H. Kauffman}
\address{University of Illinois at Chicago\\
Chicago, IL \ 60607-7045 \ \ USA}
\email{kauffman@uic.edu}
\urladdr{http://www.math.uic.edu/\symbol{126}kauffman}
\date{April 3, 2010}

\begin{abstract}
Extending the methods from our previous work on quantum knots and quantum
graphs, we describe a general procedure for quantizing a large class of
mathematical structures which includes, for example, knots, graphs, groups,
algebraic varieties, categories, topological spaces, geometric spaces, and
more. \ This procedure is different from that normally found in quantum
topology. \ We then demonstrate the power of this method by using it to
quantize braids.

This general method produces a blueprint of a quantum system which is
physically implementable in the same sense that Shor's quantum factoring
algorithm is physically implementable. \ Mathematical invariants become
objects that are physically observable.

\end{abstract}
\maketitle
\tableofcontents

\section{Introduction}

\bigskip

Extending the methods found in previous work \cite{Lomonaco1, Lomonaco2} on
quantum knots and quantum graphs, we describe a general procedure for
quantizing a large class of mathematical structures which includes, for
example, knots, graphs, groups, algebraic varieties, categories, topological
spaces, geometric spaces, and more. \ This procedure is different from that
normally found in quantum topology. \ We then demonstrate the power of this
method by using it to quantize braids.

\bigskip

We should also mention that this general method produces a blueprint of a
quantum system which is physically implementable in the same sense that Shor's
quantum factoring algorithm is physically implementable. \ Moreover,
mathematical invariants become objects that are physically observable.

\bigskip

The above mentioned general quantization procedure consists of two steps:

\begin{itemize}
\item[\textbf{Step 1.}] Mathematical construction of a motif system
$\mathcal{S}$ , and

\item[\textbf{Step 2.}] Mathematical construction of a quantum motif system
$\mathcal{Q}$ from the system $\mathcal{S}$ .
\end{itemize}

\bigskip

\noindent\textbf{Caveat.} The term "motif" used in this paper should not be
confused with the use of the term "motive" (a.k.a., "motif") found in
algebraic geometry.

\bigskip

\section{Part I. A General Procedure for Quantizing Mathematical Structures}

\bigskip

We now outline a general procedure for quantizing mathematical structures.
\ One useful advantage to this quantization procedure is that the resulting
system is a multipartite quantum system, a property that is of central
importance in quantum computation, particularly in regard to the design of
quantum algorithms. \ In a later section of this paper, we illustrate this
quantization procedure by using it to quantize braids. Examples of the
application of this quantization procedure to knots, graphs, and algebraic
structures can be found in \cite{Lomonaco1, Lomonaco2, Kauffman1}.

\bigskip

\subsection{Stage 1. \ Construction of a motif system $\mathcal{S}_{n}$}

\qquad\bigskip

Let%
\[
T=\left\{  t_{0},t_{1},\ldots,t_{\ell-1}\right\}
\]
be a finite set of symbols, with a distinguished element $t_{0}$, called the
\textbf{trivial symbol}, and with a linear ordering denoted by `$<$'. Let%
\[
T^{\times N}%
\]
be the the $N$-fold cartesian product of $T$ with an induced LEX ordering also
denoted by `$<$', and let $S\left(  n\right)  $ be the group of all
permutations of $T^{\times N}$. \ \ For positive integers $N$ and $N^{\prime}$
($N<N^{\prime}$), let
\[
\iota:T^{\times N}\longrightarrow T^{\times N^{\prime}}%
\]
be the injection defined by%
\[
\left(  t_{j(0)},t_{j(1)},t_{j(2)},\ldots,t_{j(N-1)}\right)  \longmapsto
\left(  t_{j(0)},t_{j(1)},t_{j(2)},\ldots,t_{j(N-1)},\overset{N^{\prime
}-N}{\overbrace{t_{0},t_{0},t_{0},\ldots,t_{0}}}\right)
\]

\bigskip

Next, let
\[
N_{0}<N_{1}<N_{2}<\ldots
\]
be a monotone strictly increasing infinite sequence of positive integers.

\bigskip

For each positive integer $n\geq0$, let $M^{(n)}$ be a subset of $T^{\times
N_{n}}$ such that $\iota\left(  M^{(n)}\right)  $ lies in $M^{(n+1)}$, i.e.,
$\iota\left(  M^{(n)}\right)  \subset M^{(n+1)}$. \ Moreover, for each
non-negative integer $n$, let $A(n)$ be a subgroup of the permutation group
$S\left(  N_{n}\right)  $ having $M^{(n)}$ as an invariant subset, and such
that the injection $\iota:T^{\times N_{n}}\longrightarrow T^{\times N_{n}+n}$
induces a monomorphism $\iota:A\left(  N_{n}\right)  \longrightarrow A\left(
N_{n+1}\right)  $, also denoted by $\iota$.

\bigskip

We define a \textbf{motif system} $\mathcal{S}_{n}=\mathcal{S}\left(
M^{(n)},A(n)\right)  $ of\textbf{ order} $n$ as the pair $\left(
M^{(n)},A(n)\right)  $, where $M^{(n)}$ is called the \textbf{set of motifs},
and where $A(n)$ is called the \textbf{ambient group}. \ 

\bigskip

Finally, we define a\textbf{ nested motif system} $\mathcal{S}_{\ast
}=\mathcal{S}_{\ast}\left(  M^{(\ast)},A(\ast)\right)  $ as the following
sequence of sets, groups, injections, and monomorphisms:%
\[
\mathcal{S}_{1}\left(  M^{(1)},A(1)\right)  \overset{\iota}{\longrightarrow
}\mathcal{S}_{2}\left(  M^{(2)},A(2)\right)  \overset{\iota}{\longrightarrow
}\cdots\overset{\iota}{\longrightarrow}\mathcal{S}_{n}\left(  M^{(n)}%
,A(n)\right)  \overset{\iota}{\longrightarrow}\cdots
\]

\bigskip

\begin{remark}
There is also one more symbolic motif system that is often of use, the
\textbf{direct limit motif system} defined by%
\[
\mathcal{S}_{\infty}\left(  M^{(\infty)},A(\infty)\right)  =\lim
_{\longrightarrow}\mathcal{S}_{\ast}\left(  M^{(\ast)},A(\ast)\right)  \text{
,}%
\]
where $\lim\limits_{\longrightarrow}$ denotes the direct limit.
\end{remark}

\bigskip

\subsection{Stage 2. Motif equivalence and motif invariants}

\qquad\bigskip

Let $\mathcal{S}_{n}=\mathcal{S}\left(  M^{(n)},A(n)\right)  $ be a motif
system of order $n$.

\bigskip

Two motifs $m_{1}$ and $m_{2}$ of the set $\mathcal{M}^{(n)}$ are said to be
of the \textbf{same }$n$\textbf{-motif type}, written%
\[
m_{1}\underset{n}{\sim}m_{2}\text{ ,}%
\]
if here exists an element $g$ of the ambient group $A(n)$ which takes $m_{1}$
to $m_{2}$, i.e., such that%
\[
gm_{1}=m_{2}\text{ .}%
\]
The motifs $m_{1}$ and $m_{2}$ are said to be of the \textbf{same motif type},
written%
\[
m_{1}\sim m_{2}\text{ ,}%
\]
if there exists a non-negative integer $k$ such that%
\[
\iota^{k}m_{1}\underset{n+k}{\sim}\iota^{k}m_{2}\text{ .}%
\]

\bigskip

We now wish to answer the question:

\bigskip

\noindent\textbf{Question}. \ \textit{What is meant by a motif invariant?}

\bigskip

\begin{definition}
Let $\mathcal{S}_{n}=\mathcal{S}\left(  M^{(n)},A(n)\right)  $ be a motif
system, and let $\mathbb{D}$ be some yet to be chosen mathematical domain.
\ By an $n$-\textbf{motif invariant} $I^{(n)}$, we mean a map
\[
I^{(n)}:M^{(n)}\longrightarrow\mathbb{D}%
\]
such that, when two motifs $m_{1}$ and $m_{2}$ are of the same $n$-type,
i.e.,
\[
m_{1}\underset{n}{\sim}m_{2}\text{ , }%
\]
then their respective invariants must be equal, i.e.,
\[
I^{(n)}\left(  m_{1}\right)  =I^{(n)}\left(  m_{2}\right)  \text{ .}%
\]
In other words, $I^{(n)}:M^{(n)}\longrightarrow\mathbb{D}$ is a map that is
invariant under the action of the ambient group $A(n)$, i.e.,
\[
I^{(n)}\left(  m\right)  =I^{(n)}\left(  gm\right)
\]
for all elements of $g$ in $A(n)$.
\end{definition}

\bigskip

\subsection{Stage 3. Construction of the corresponding quantum motif systems
$\mathcal{Q}_{n}$}

\quad

\bigskip

We now use the nested motif system $\mathcal{S}_{\ast}$ to construct a nested
sequence of quantum motif systems $\mathcal{Q}_{\ast}$.

\bigskip

For each non-negative $n,$ the corresponding $n$\textbf{-th order}
\textbf{quantum motif system}%
\[
\mathcal{Q}_{n}=\mathcal{Q}\left(  \mathcal{M}^{(n)},\mathcal{A}(n)\right)
\]
consists of a Hilbert space $\mathcal{M}^{(n)}$, called the \textbf{quantum
motif space}, and a group $\mathcal{A}(n)$, also called the \textbf{ambient
group}.\ The quantum motif space $\mathcal{M}^{(n)}$ and the ambient group
$\mathcal{A}(n)$ are defined as follows:\bigskip

\begin{itemize}
\item The\textbf{ quantum motif space} $\mathcal{M}^{(n)}$ is the Hilbert
space with orthonormal basis%
\[
\left\{  \ \left\vert m\right\rangle :m\in M^{(n)}\ \right\}  \text{ .}%
\]
The elements of $\mathcal{M}^{(n)}$ are called \textbf{quantum motifs}%
$.$\bigskip

\item The \textbf{ambient group} $\mathcal{A}(n)$ is the unitary group acting
on the Hilbert space $\mathcal{M}^{(n)}$ consisting of all linear
transformations of the form%
\[
\left\{  \ \widetilde{g}:\mathcal{M}^{(n)}\longrightarrow\mathcal{M}%
^{(n)}:g\in A(n)\ \right\}  \text{ ,}%
\]
where $\widetilde{g}$ is the linear transformation defined by
\[%
\begin{array}
[c]{rcc}%
\widetilde{g}:\mathcal{M}^{(n)} & \longrightarrow & \mathcal{M}^{(n)}\\
\left\vert m\right\rangle \quad & \longmapsto & \left\vert gm\right\rangle
\end{array}
\]
Since each element $g$ in $A(n)$ is a permutation, each $\widetilde{g}$
permutes the orthonormal basis $\left\{  \ \left\vert m\right\rangle :m\in
M^{(n)}\ \right\}  $ of $\mathcal{M}^{(n)}$. \ Hence, $\widetilde{g}$ is
automatically a unitary transformation. \ It follows that $A(n)$ and
$\mathcal{A}(n)$ are isomorphic as groups. \ We will often abuse notation by
denoting $\widetilde{g}$ by $g$, and $\mathcal{A}(n)$ by $A(n)$.
\end{itemize}

\bigskip

Next, for each non-negative integer $n$, let
\[
\iota:\mathcal{M}^{(n)}\longrightarrow\mathcal{M}^{(n+1)}%
\]
and%
\[
\iota:\mathcal{A}(n)\longrightarrow\mathcal{A}(n+1)
\]
respectively denote the Hilbert space monomorphism and the group monomorphism
induced by the injection%
\[
\iota:M^{(n)}\longrightarrow M^{(n+1)}%
\]
and the group monomorphism%
\[
\iota:A(n)\longrightarrow A(n+1)\text{ .}%
\]

\bigskip

Finally, we define the\textbf{ nested quantum motif system} $\mathcal{Q}%
_{\ast}=\mathcal{Q}_{\ast}\left(  \mathcal{M}^{(\ast)},\mathcal{A}%
(\ast)\right)  $ as the following sequence of Hilbert spaces, groups, Hilbert
space monomorphisms, and group monomorphisms:%
\[
\mathcal{Q}_{1}\left(  \mathcal{M}^{(1)},\mathcal{A}(1)\right)  \overset{\iota
}{\longrightarrow}\mathcal{Q}_{2}\left(  \mathcal{M}^{(2)},\mathcal{A}%
(2)\right)  \overset{\iota}{\longrightarrow}\cdots\overset{\iota
}{\longrightarrow}\mathcal{Q}_{n}\left(  \mathcal{M}^{(n)},\mathcal{A}%
(n)\right)  \overset{\iota}{\longrightarrow}\cdots
\]

\bigskip

\begin{remark}
We should also mention one other quantum motif system that can be useful,
namely, the\textbf{ quantum direct limit motif system} defined by%
\[
\mathcal{Q}_{\infty}=\mathcal{Q}_{\infty}\left(  \mathcal{M}^{(\infty
)},\mathcal{A}(\infty)\right)  =\lim_{\longrightarrow}\mathcal{Q}_{\ast
}\left(  \mathcal{M}^{(\ast)},\mathcal{A}(\ast)\right)  \text{ ,}%
\]
where $\lim\limits_{\longrightarrow}$ denotes the direct limit. \ This quantum
system is often also physically implementable.
\end{remark}

\bigskip

\subsection{Stage 4. Quantum motif equivalence}

\bigskip

Let $\mathcal{Q}_{n}=\mathcal{Q}\left(  \mathcal{M}^{(n)},\mathcal{A}%
(n)\right)  $ be a quantum motif system of order $n$.

\bigskip

Two quantum motifs $\left\vert \psi_{1}\right\rangle $ and $\left\vert
\psi_{2}\right\rangle $ of the Hilbert space $\mathcal{M}^{(n)}$ are said to
be of the \textbf{same }$n$\textbf{-motif type}, written%
\[
\left\vert \psi_{1}\right\rangle \underset{n}{\sim}\left\vert \psi
_{2}\right\rangle \text{ ,}%
\]
if there exists an element $g$ of the ambient group $\mathcal{A}(n)$ which
takes $\left\vert \psi_{1}\right\rangle $ to $\left\vert \psi_{2}\right\rangle
$, i.e., such that%
\[
g\left\vert \psi_{1}\right\rangle =\left\vert \psi_{2}\right\rangle \text{ .}%
\]
The quantum motifs $\left\vert \psi_{1}\right\rangle $ and $\left\vert
\psi_{2}\right\rangle $ are said to be of the \textbf{same motif type},
written%
\[
\left\vert \psi_{1}\right\rangle \sim\left\vert \psi_{2}\right\rangle \text{
,}%
\]
if there exists a non-negative integer $m$ such that%
\[
\iota^{m}\left\vert \psi_{1}\right\rangle \underset{n+m}{\sim}\iota
^{m}\left\vert \psi_{2}\right\rangle \text{ .}%
\]

\bigskip

\subsection{Stage 5. Motif invariants as quantum observables}

\qquad\bigskip

We consider the following question:

\bigskip

\noindent\textbf{Question:} \ \textit{What do we mean by a physically
observable quantum motif invariant?}

\bigskip

We answer this question with a definition.

\bigskip

\begin{definition}
Let $\mathcal{Q}_{n}=\mathcal{Q}\left(  \mathcal{M}^{(n)},\mathcal{A}%
(n)\right)  $ be a quantum motif system of order $n$, and let $\Omega$ be an
observable, i.e., a Hermitian operator on the Hilbert space $\mathcal{M}%
^{(n)}$ of quantum motifs. \ Then $\Omega$ is a \textbf{quantum motif }%
$n$\textbf{-invariant} provided $\Omega$ is left invariant under the big
adjoint action of the ambient group $\mathcal{A}(n)$, i.e., provided
\[
U\Omega U^{-1}=\Omega
\]
for all $U$ in $\mathcal{A}(n)$.
\end{definition}

\bigskip

\begin{proposition}
If%
\[
I^{(n)}:M^{(n)}\longrightarrow\mathbb{R}%
\]
is a real valued $n$-motif invariant, then%
\[
\Omega=%
{\displaystyle\sum\limits_{m\in M^{(n)}}}
I^{(n)}\left(  m\right)  \left\vert m\right\rangle \left\langle m\right\vert
\]
is a quantum motif observable which is a quantum motif $n$-invariant.
\end{proposition}

\bigskip

Much more can be said about this topic. \ \ For a more in-depth discussion of
this issue, we refer the reader to \cite{Lomonaco1, Lomonaco2}.

\section{Part II. Quantizing Braids}

\bigskip

We now illustrate the quantization procedure defined above by using it to
quantize braids.

\bigskip

\subsection{Stage 1. The set of braid mosaics $\mathbb{B}^{(n,\ell)}$}

\qquad\bigskip

For each integer $n\geq2$, let $\mathbb{T}^{(n)}$ denote the following set of
the $2n-1$ symbols%
\[%
{\includegraphics[
natheight=6.000100in,
natwidth=3.000000in,
height=0.627in,
width=0.3269in
]%
{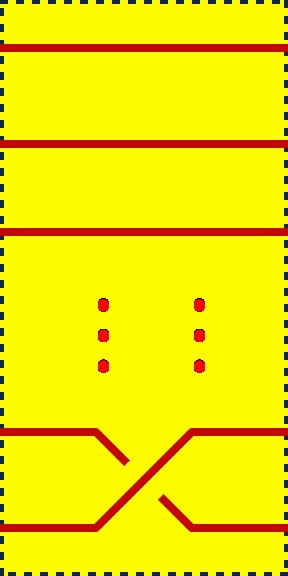}%
}
,\cdots\text{,}\
{\includegraphics[
natheight=6.000100in,
natwidth=3.000000in,
height=0.627in,
width=0.3269in
]%
{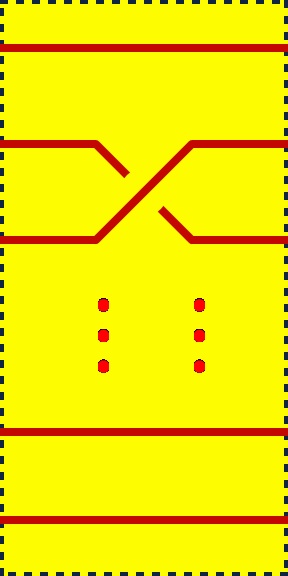}%
}
,\
{\includegraphics[
natheight=6.000100in,
natwidth=3.000000in,
height=0.627in,
width=0.3269in
]%
{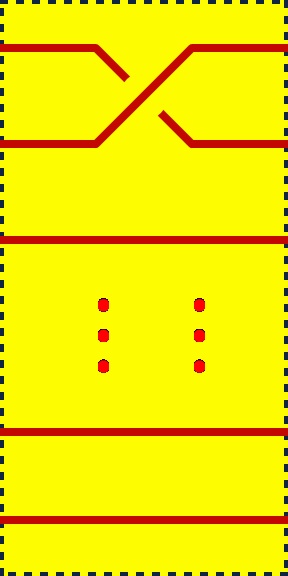}%
}
,\
{\includegraphics[
natheight=6.000100in,
natwidth=3.000000in,
height=0.627in,
width=0.3269in
]%
{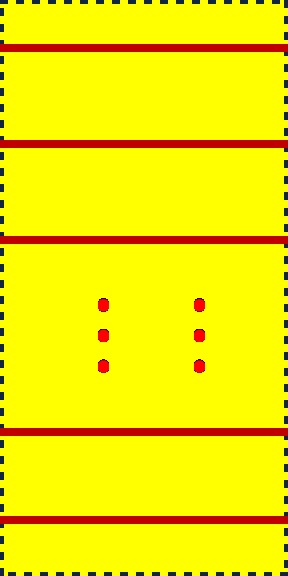}%
}
,\
{\includegraphics[
natheight=6.000100in,
natwidth=3.000000in,
height=0.627in,
width=0.3269in
]%
{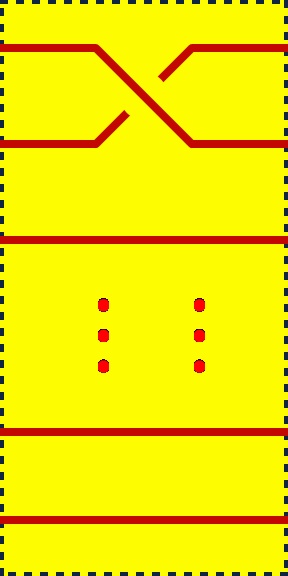}%
}
,\
{\includegraphics[
natheight=6.000100in,
natwidth=3.000000in,
height=0.627in,
width=0.3269in
]%
{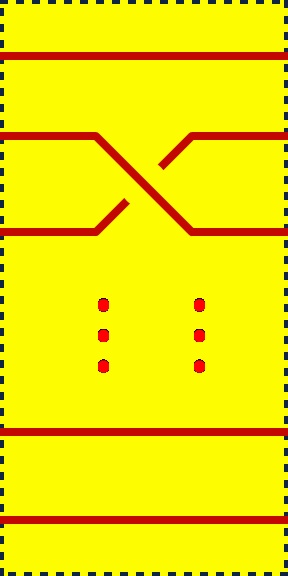}%
}
,\cdots\text{,}\
{\includegraphics[
natheight=6.000100in,
natwidth=3.000000in,
height=0.627in,
width=0.3269in
]%
{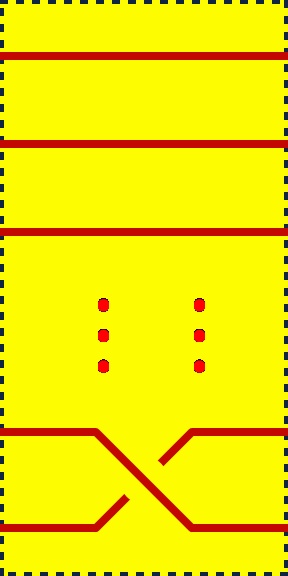}%
}
\]
called $\mathbf{n}$-\textbf{stranded braid tiles}, or $n$-\textbf{tiles}, or
simply \textbf{tiles}. \ We also denote these tiles respectively by the
symbols
\[
b_{-(n-1)},\ldots,b_{-2},b_{-1},b_{0}=1,b_{1},b_{2},\ldots,b_{n-1}\text{ ,}%
\]
as indicated in the table given below:%
\[%
\begin{tabular}
[c]{c}%
\begin{tabular}
[c]{||c||c||c||c||c||c||c||c||c||}\hline\hline
$%
{\includegraphics[
natheight=6.000100in,
natwidth=3.000000in,
height=0.627in,
width=0.3269in
]%
{rtmmn.jpg}%
}
$ & $\cdots$ & $%
{\includegraphics[
natheight=6.000100in,
natwidth=3.000000in,
height=0.627in,
width=0.3269in
]%
{rtm2n.jpg}%
}
$ & $%
{\includegraphics[
natheight=6.000100in,
natwidth=3.000000in,
height=0.627in,
width=0.3269in
]%
{rtm1n.jpg}%
}
$ & $\overset{}{%
{\includegraphics[
natheight=6.000100in,
natwidth=3.000000in,
height=0.627in,
width=0.3269in
]%
{rt0n.jpg}%
}
}$ & $%
{\includegraphics[
natheight=6.000100in,
natwidth=3.000000in,
height=0.627in,
width=0.3269in
]%
{rtp1n.jpg}%
}
$ & $%
{\includegraphics[
natheight=6.000100in,
natwidth=3.000000in,
height=0.627in,
width=0.3269in
]%
{rtp2n.jpg}%
}
$ & $\cdots$ & $%
{\includegraphics[
natheight=6.000100in,
natwidth=3.000000in,
height=0.627in,
width=0.3269in
]%
{rtpmn.jpg}%
}
$\\\hline
$b_{-(n-1)}$ & $\cdots$ & $b_{-2}$ & $b_{-1}$ & $b_{0}=1$ &
$\overset{}{\underset{}{b_{1}}}$ & $b_{2}$ & $\cdots$ & $b_{n-1}%
$\\\hline\hline
\end{tabular}
\\
$n$\textbf{-stranded braid tiles}%
\end{tabular}
\ \
\]

\bigskip

\begin{definition}
An $\left(  \mathbf{n},\mathbf{\ell}\right)  $\textbf{-braid mosaic} $\beta$
is defined as a sequence of $n$-stranded braid tiles%
\[
\beta=b_{j(1)}b_{j(2)}\ldots b_{j(\ell)}%
\]
of length $\ell$. \ We let $\mathbb{B}^{(n,\ell)}$ denote the \textbf{set of
all }$\left(  n,\mathbf{\ell}\right)  $\textbf{-braid mosaics}.
\end{definition}

\bigskip

An example of a $(3,8)$-braid mosaic is given below%

\[%
\begin{tabular}
[c]{c}%
$%
\begin{array}
[c]{cccccccc}%
\hspace{-0.1in}%
{\includegraphics[
natheight=3.000000in,
natwidth=3.000000in,
height=0.3269in,
width=0.3269in
]%
{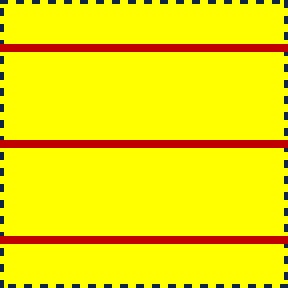}%
}
& \hspace{-0.1in}%
{\includegraphics[
natheight=3.000000in,
natwidth=3.000000in,
height=0.3269in,
width=0.3269in
]%
{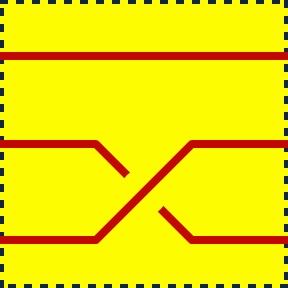}%
}
& \hspace{-0.1in}%
{\includegraphics[
natheight=3.000000in,
natwidth=3.000000in,
height=0.3269in,
width=0.3269in
]%
{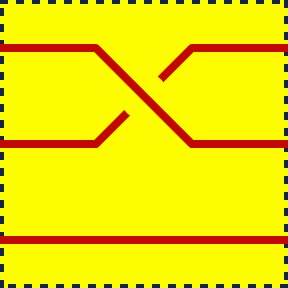}%
}
& \hspace{-0.1in}%
{\includegraphics[
natheight=3.000000in,
natwidth=3.000000in,
height=0.3269in,
width=0.3269in
]%
{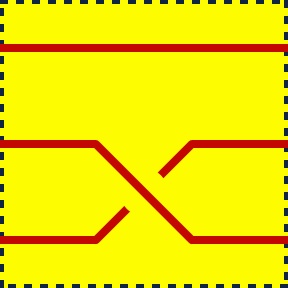}%
}
& \hspace{-0.1in}%
{\includegraphics[
natheight=3.000000in,
natwidth=3.000000in,
height=0.3269in,
width=0.3269in
]%
{rt03.jpg}%
}
& \hspace{-0.1in}%
{\includegraphics[
natheight=3.000000in,
natwidth=3.000000in,
height=0.3269in,
width=0.3269in
]%
{rt03.jpg}%
}
& \hspace{-0.1in}%
{\includegraphics[
natheight=3.000000in,
natwidth=3.000000in,
height=0.3269in,
width=0.3269in
]%
{rtm23.jpg}%
}
& \hspace{-0.1in}%
{\includegraphics[
natheight=3.000000in,
natwidth=3.000000in,
height=0.3269in,
width=0.3269in
]%
{rtp23.jpg}%
}
\\
1 & b_{-1} & b_{1} & b_{2} & 1 & 1 & b_{-1} & b_{2}%
\end{array}
$\\
The $\left(  3,8\right)  $-braid mosaic $\beta=1b_{-1}b_{1}b_{2}11b_{-1}b_{2}$%
\end{tabular}
\ \ \
\]

\bigskip

\begin{remark}
Please note that the set of all $\left(  n,\ell\right)  $-braid mosaics
$\mathbb{B}^{(n,\ell)}$ is a finite set of cardinality $\left(  2n-1\right)
^{\ell}$.
\end{remark}

\bigskip

\subsection{Stage 1 (Cont.) Braid mosaic moves}

\bigskip

\begin{definition}
Let $\ell^{\prime}$ and $\ell$ be positive integers such that $\ell^{\prime
}\leq\ell$. \ An $\left(  n,\ell^{\prime}\right)  $-braid mosaic $\gamma$ is
is said to be an $\left(  n,\ell^{\prime}\right)  $\textbf{-braid submosaic}
of an $\left(  n,\ell\right)  $-braid mosaic $\beta$ provided $\gamma$ is a
subsequence of consecutive tiles of $\beta$. \ The $\left(  n,\ell^{\prime
}\right)  $-braid submosaic $\gamma$ is said to be at \textbf{position} $p$ in
$\beta$ if the first (leftmost) tile of $\gamma$ is the $p$-th tile of $\beta$
from the left. \ We denote the $\left(  n,\ell^{\prime}\right)  $-braid
submosaic $\gamma$ of $\beta$ at location $p$ by $\gamma=\beta^{p:\ell
^{\prime}}$.
\end{definition}

\bigskip

\begin{remark}
The number of $\left(  n,\ell^{\prime}\right)  $-braid submosaics of an
$\left(  n,\ell\right)  $-braid mosaic $\beta$ is $\ell-\ell^{\prime}+1$.
\end{remark}

\bigskip

Two examples of braid submosaics of the $\left(  3,8\right)  $-braid mosaic
$\beta=1b_{-1}b_{1}b_{2}11b_{-1}b_{2}$ are given above are:%
\[%
\begin{tabular}
[c]{c}%
$%
\begin{array}
[c]{ccc}%
\hspace{-0.1in}%
{\includegraphics[
natheight=3.000000in,
natwidth=3.000000in,
height=0.3269in,
width=0.3269in
]%
{rtm23.jpg}%
}
& \hspace{-0.1in}%
{\includegraphics[
natheight=3.000000in,
natwidth=3.000000in,
height=0.3269in,
width=0.3269in
]%
{rtp13.jpg}%
}
& \hspace{-0.1in}%
{\includegraphics[
natheight=3.000000in,
natwidth=3.000000in,
height=0.3269in,
width=0.3269in
]%
{rtp23.jpg}%
}
\\
b_{-1} & b_{1} & b_{2}%
\end{array}
$\\
\textbf{The }$\left(  3,3\right)  $\textbf{-braid submosaic}\\
$\beta^{2:3}$ \textbf{of }$\beta$\textbf{ at position 2}%
\end{tabular}
\ \ \ \ \text{ \ \ \ \ \ \ \ \ \ \ \ \ \ \ \ \ \ }%
\begin{tabular}
[c]{c}%
$%
\begin{array}
[c]{cccc}%
\hspace{-0.1in}%
{\includegraphics[
natheight=3.000000in,
natwidth=3.000000in,
height=0.3269in,
width=0.3269in
]%
{rt03.jpg}%
}
& \hspace{-0.1in}%
{\includegraphics[
natheight=3.000000in,
natwidth=3.000000in,
height=0.3269in,
width=0.3269in
]%
{rt03.jpg}%
}
& \hspace{-0.1in}%
{\includegraphics[
natheight=3.000000in,
natwidth=3.000000in,
height=0.3269in,
width=0.3269in
]%
{rtm23.jpg}%
}
& \hspace{-0.1in}%
{\includegraphics[
natheight=3.000000in,
natwidth=3.000000in,
height=0.3269in,
width=0.3269in
]%
{rtp23.jpg}%
}
\\
1 & 1 & b_{-1} & b_{2}%
\end{array}
$\\
\textbf{The }$\left(  3,4\right)  $\textbf{-braid submosaic}\\
$\beta^{5:4}$ \textbf{of }$\beta$\textbf{ at position 5}%
\end{tabular}
\ \ \ \
\]

\bigskip

\begin{definition}
Let $\ell^{\prime}$ and $\ell$ be positive integers such that $\ell^{\prime
}\leq\ell$. \ For any two $\left(  n,\ell^{\prime}\right)  $-braid mosaics
$\gamma$ and $\gamma^{\prime}$, we define the $\ell^{\prime}$-\textbf{braid
mosaic move} \textbf{at location} $p$ on the set of all $\left(
n,\ell\right)  $-braid mosaics $\mathbb{B}^{\left(  n,\ell\right)  }$, denoted
by%
\[
\gamma\overset{p}{\leftrightarrow}\gamma^{\prime}\text{ ,}%
\]
as the map defined by%
\[
\left(  \gamma\overset{p}{\leftrightarrow}\gamma^{\prime}\right)  \left(
\beta\right)  =\left\{
\begin{array}
[c]{ll}%
\beta\text{ with }\beta^{p:\ell^{\prime}}\text{ replaced by }\gamma^{\prime} &
\text{if }\beta^{p:\ell^{\prime}}=\gamma\\
& \\
\beta\text{ with }\beta^{p:\ell^{\prime}}\text{ replaced by }\gamma & \text{if
}\beta^{p:\ell^{\prime}}=\gamma^{\prime}\\
& \\
\beta & \text{otherwise}%
\end{array}
\right\vert
\]

\end{definition}

\bigskip

As an example, consider the $2$-braid mosaic move $\gamma
\overset{3}{\leftrightarrow}\gamma^{\prime}$ at position $3$ defined by%
\[
\gamma\overset{3}{\leftrightarrow}\gamma^{\prime}\qquad=\qquad%
\begin{array}
[c]{cc}%
{\includegraphics[
natheight=3.000000in,
natwidth=3.000000in,
height=0.3269in,
width=0.3269in
]%
{rtm23.jpg}%
}
& \hspace{-0.1in}%
{\includegraphics[
natheight=3.000000in,
natwidth=3.000000in,
height=0.3269in,
width=0.3269in
]%
{rtp13.jpg}%
}
\end{array}
\overset{3}{\longleftrightarrow}%
\begin{array}
[c]{cc}%
{\includegraphics[
natheight=3.000000in,
natwidth=3.000000in,
height=0.3269in,
width=0.3269in
]%
{rt03.jpg}%
}
& \hspace{-0.1in}%
{\includegraphics[
natheight=3.000000in,
natwidth=3.000000in,
height=0.3269in,
width=0.3269in
]%
{rtm23.jpg}%
}
\end{array}
\]
Then%
\[
\hspace{-0.75in}\left(  \gamma\overset{3}{\leftrightarrow}\gamma^{\prime
}\right)  \left(
\begin{array}
[c]{ccccc}%
\hspace{-0.1in}%
{\includegraphics[
natheight=3.000000in,
natwidth=3.000000in,
height=0.3269in,
width=0.3269in
]%
{rt03.jpg}%
}
& \hspace{-0.1in}%
{\includegraphics[
natheight=3.000000in,
natwidth=3.000000in,
height=0.3269in,
width=0.3269in
]%
{rtp23.jpg}%
}
& \hspace{-0.1in}%
{\includegraphics[
natheight=3.000000in,
natwidth=3.000000in,
height=0.3269in,
width=0.3269in
]%
{rtm23.jpg}%
}
& \hspace{-0.1in}%
{\includegraphics[
natheight=3.000000in,
natwidth=3.000000in,
height=0.3269in,
width=0.3269in
]%
{rtp13.jpg}%
}
& \hspace{-0.1in}%
{\includegraphics[
natheight=3.000000in,
natwidth=3.000000in,
height=0.3269in,
width=0.3269in
]%
{rtp23.jpg}%
}
\end{array}
\right)  =\
\begin{array}
[c]{ccccc}%
\hspace{-0.1in}%
{\includegraphics[
natheight=3.000000in,
natwidth=3.000000in,
height=0.3269in,
width=0.3269in
]%
{rt03.jpg}%
}
& \hspace{-0.1in}%
{\includegraphics[
natheight=3.000000in,
natwidth=3.000000in,
height=0.3269in,
width=0.3269in
]%
{rtp23.jpg}%
}
& \hspace{-0.1in}%
{\includegraphics[
natheight=3.000000in,
natwidth=3.000000in,
height=0.3269in,
width=0.3269in
]%
{rt03.jpg}%
}
& \hspace{-0.1in}%
{\includegraphics[
natheight=3.000000in,
natwidth=3.000000in,
height=0.3269in,
width=0.3269in
]%
{rtm23.jpg}%
}
& \hspace{-0.1in}%
{\includegraphics[
natheight=3.000000in,
natwidth=3.000000in,
height=0.3269in,
width=0.3269in
]%
{rtp23.jpg}%
}
\end{array}
\ \ \left(
\begin{tabular}
[c]{c}%
Braid\\
Submosics\\
Switched
\end{tabular}
\ \ \ \ \ \right)
\]%
\[
\hspace{-0.75in}\left(  \gamma\overset{3}{\leftrightarrow}\gamma^{\prime
}\right)  \left(
\begin{array}
[c]{ccccc}%
\hspace{-0.1in}%
{\includegraphics[
natheight=3.000000in,
natwidth=3.000000in,
height=0.3269in,
width=0.3269in
]%
{rt03.jpg}%
}
& \hspace{-0.1in}%
{\includegraphics[
natheight=3.000000in,
natwidth=3.000000in,
height=0.3269in,
width=0.3269in
]%
{rtp23.jpg}%
}
& \hspace{-0.1in}%
{\includegraphics[
natheight=3.000000in,
natwidth=3.000000in,
height=0.3269in,
width=0.3269in
]%
{rt03.jpg}%
}
& \hspace{-0.1in}%
{\includegraphics[
natheight=3.000000in,
natwidth=3.000000in,
height=0.3269in,
width=0.3269in
]%
{rtm23.jpg}%
}
& \hspace{-0.1in}%
{\includegraphics[
natheight=3.000000in,
natwidth=3.000000in,
height=0.3269in,
width=0.3269in
]%
{rtp23.jpg}%
}
\end{array}
\right)  =\
\begin{array}
[c]{ccccc}%
\hspace{-0.1in}%
{\includegraphics[
natheight=3.000000in,
natwidth=3.000000in,
height=0.3269in,
width=0.3269in
]%
{rt03.jpg}%
}
& \hspace{-0.1in}%
{\includegraphics[
natheight=3.000000in,
natwidth=3.000000in,
height=0.3269in,
width=0.3269in
]%
{rtp23.jpg}%
}
& \hspace{-0.1in}%
{\includegraphics[
natheight=3.000000in,
natwidth=3.000000in,
height=0.3269in,
width=0.3269in
]%
{rtm23.jpg}%
}
& \hspace{-0.1in}%
{\includegraphics[
natheight=3.000000in,
natwidth=3.000000in,
height=0.3269in,
width=0.3269in
]%
{rtp13.jpg}%
}
& \hspace{-0.1in}%
{\includegraphics[
natheight=3.000000in,
natwidth=3.000000in,
height=0.3269in,
width=0.3269in
]%
{rtp23.jpg}%
}
\end{array}
\ \ \left(
\begin{tabular}
[c]{c}%
Braid\\
Submosics\\
Switched
\end{tabular}
\ \ \ \ \ \right)
\]%
\[
\hspace{-0.75in}\left(  \gamma\overset{3}{\leftrightarrow}\gamma^{\prime
}\right)  \left(
\begin{array}
[c]{ccccc}%
\hspace{-0.1in}%
{\includegraphics[
natheight=3.000000in,
natwidth=3.000000in,
height=0.3269in,
width=0.3269in
]%
{rt03.jpg}%
}
& \hspace{-0.1in}%
{\includegraphics[
natheight=3.000000in,
natwidth=3.000000in,
height=0.3269in,
width=0.3269in
]%
{rtp23.jpg}%
}
& \hspace{-0.1in}%
{\includegraphics[
natheight=3.000000in,
natwidth=3.000000in,
height=0.3269in,
width=0.3269in
]%
{rtp13.jpg}%
}
& \hspace{-0.1in}%
{\includegraphics[
natheight=3.000000in,
natwidth=3.000000in,
height=0.3269in,
width=0.3269in
]%
{rtm23.jpg}%
}
& \hspace{-0.1in}%
{\includegraphics[
natheight=3.000000in,
natwidth=3.000000in,
height=0.3269in,
width=0.3269in
]%
{rtp13.jpg}%
}
\end{array}
\right)  =\
\begin{array}
[c]{ccccc}%
\hspace{-0.1in}%
{\includegraphics[
natheight=3.000000in,
natwidth=3.000000in,
height=0.3269in,
width=0.3269in
]%
{rt03.jpg}%
}
& \hspace{-0.1in}%
{\includegraphics[
natheight=3.000000in,
natwidth=3.000000in,
height=0.3269in,
width=0.3269in
]%
{rtp23.jpg}%
}
& \hspace{-0.1in}%
{\includegraphics[
natheight=3.000000in,
natwidth=3.000000in,
height=0.3269in,
width=0.3269in
]%
{rtp13.jpg}%
}
& \hspace{-0.1in}%
{\includegraphics[
natheight=3.000000in,
natwidth=3.000000in,
height=0.3269in,
width=0.3269in
]%
{rtm23.jpg}%
}
& \hspace{-0.1in}%
{\includegraphics[
natheight=3.000000in,
natwidth=3.000000in,
height=0.3269in,
width=0.3269in
]%
{rtp13.jpg}%
}
\end{array}
\ \ \left(
\begin{tabular}
[c]{c}%
Braid\\
Mosaic\\
Unchanged
\end{tabular}
\ \ \ \ \ \right)
\]

\bigskip

The following proposition is an almost immediate consequence of the definition
of a braid move.

\bigskip

\begin{proposition}
Each braid move is a permutation on the set $\mathbb{B}^{(n,\ell)}$ of
$(n,\ell)$-braid mosaics. \ In fact, it is a permutation which is a product of
disjoint transpositions.
\end{proposition}

\bigskip

\subsection{Stage 1. (Cont.) Planar isotopy moves}

\qquad\bigskip

Our next objective is to translate all the standard topological moves on
braids into braid mosaic moves. \ To accomplish this, we must first note that
there are two types of standard topological moves, i.e., those which do not
change the topological type of the braid projection, called \textbf{planar
isotopy moves}, and those which do change the typological type of the braid
projection but not of the braid itself, called \textbf{Reidemeister moves}.

\bigskip

We begin with the planar isotopy moves.

\bigskip

\begin{definition}
For braid mosaics, there are two types \textbf{planar isotopy moves}, i.e.,
types $P_{1}$ and $P_{2}$, which are defined below as:%
\[%
\begin{tabular}
[c]{|c|}\hline
$%
\begin{array}
[c]{c}%
\ \\
\
\end{array}
1b_{i}\overset{\lambda}{\underset{P_{1}}{\longleftrightarrow}}b_{i}1\text{
\ for \ }0<\left\vert i\right\vert <n%
\begin{array}
[c]{c}%
\ \\
\
\end{array}
$\\\hline
\textbf{Definition of a type }$P_{1}$\textbf{ planar isotopy move}\\\hline
\end{tabular}
\]
and%
\[%
\begin{tabular}
[c]{|c|}\hline
$%
\begin{array}
[c]{c}%
\ \\
\
\end{array}
b_{i}b_{j}\overset{\lambda}{\underset{P_{2}}{\longleftrightarrow}}b_{i}%
b_{j}\ for0<\left\vert i\right\vert ,\left\vert j\right\vert
<n\ and\ \left\vert \left\vert i\right\vert -\left\vert j\right\vert
\right\vert >1%
\begin{array}
[c]{c}%
\ \\
\
\end{array}
$\\\hline
\textbf{Definition of a type }$P_{2}$\textbf{ planar isotopy move}\\\hline
\end{tabular}
\]
\ 
\end{definition}

\bigskip

\begin{example}
Examples of $P_{1}$ and $P_{2}$ moves are respectively given below:%
\[%
\begin{tabular}
[c]{c}%
{\includegraphics[
natheight=3.000000in,
natwidth=3.000000in,
height=0.3269in,
width=0.3269in
]%
{rt03.jpg}%
}
{\includegraphics[
natheight=3.000000in,
natwidth=3.000000in,
height=0.3269in,
width=0.3269in
]%
{rtp13.jpg}%
}
$\overset{\lambda}{\longleftrightarrow}$%
{\includegraphics[
natheight=3.000000in,
natwidth=3.000000in,
height=0.3269in,
width=0.3269in
]%
{rtp13.jpg}%
}
{\includegraphics[
natheight=3.000000in,
natwidth=3.000000in,
height=0.3269in,
width=0.3269in
]%
{rt03.jpg}%
}
\\
$\text{An example of a }P_{1}\text{ move.}$%
\end{tabular}
\ \text{ \ \ and \ \ }%
\begin{tabular}
[c]{c}%
{\includegraphics[
natheight=3.999800in,
natwidth=3.000000in,
height=0.4272in,
width=0.3269in
]%
{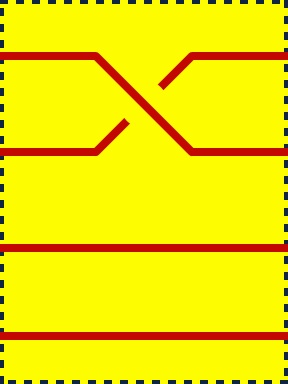}%
}
{\includegraphics[
natheight=3.999800in,
natwidth=3.000000in,
height=0.4272in,
width=0.3269in
]%
{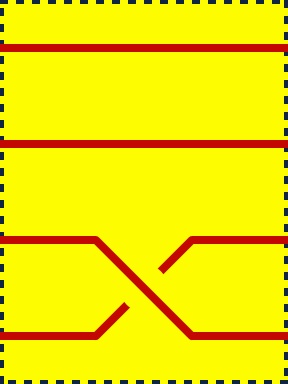}%
}
$\overset{\lambda}{\longleftrightarrow}$%
{\includegraphics[
natheight=3.999800in,
natwidth=3.000000in,
height=0.4272in,
width=0.3269in
]%
{rtp34.jpg}%
}
{\includegraphics[
natheight=3.999800in,
natwidth=3.000000in,
height=0.4272in,
width=0.3269in
]%
{rtp14.jpg}%
}
\\
$\text{An example of a }P_{2}\text{ move: }$%
\end{tabular}
\
\]

\end{example}

\bigskip

\begin{remark}
The number of $P_{1}$ and $P_{2}$ moves are respectively $2\left(  n-1\right)
\left(  \ell-1\right)  $ \ and \ $\left(  n-1\right)  \left(  2n-6\right)
\left(  \ell-1\right)  $ .
\end{remark}

\bigskip

\subsection{Stage 1. (Cont.) Reidemeister moves}

\qquad\bigskip

There are two types of topological moves, i.e., $R_{2}$ and $R_{3}$.

\bigskip

\begin{definition}
The \textbf{Reidemeister} $R_{2}$ moves are defined as%
\[%
\begin{tabular}
[c]{c}%
$b_{i}b_{-i}\overset{\lambda}{\longleftrightarrow}1^{2}$\\
where$\text{ \ }0<\left\vert i\right\vert <n$%
\end{tabular}
\]

\end{definition}

\bigskip

\begin{example}
An example of a Reidemeister 2 move is given below%
\[%
{\includegraphics[
natheight=3.000000in,
natwidth=3.000000in,
height=0.3269in,
width=0.3269in
]%
{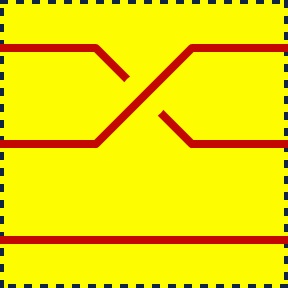}%
}
{\includegraphics[
natheight=3.000000in,
natwidth=3.000000in,
height=0.3269in,
width=0.3269in
]%
{rtp13.jpg}%
}
\overset{\lambda}{\longleftrightarrow}%
{\includegraphics[
natheight=3.000000in,
natwidth=3.000000in,
height=0.3269in,
width=0.3269in
]%
{rt03.jpg}%
}
{\includegraphics[
natheight=3.000000in,
natwidth=3.000000in,
height=0.3269in,
width=0.3269in
]%
{rt03.jpg}%
}
\]

\end{example}

\bigskip

\begin{remark}
The number of $R_{2}$ moves is $2(n-1)\left(  \ell-1\right)  $
\end{remark}

\bigskip

\begin{definition}
The \textbf{Reidemeister} $R_{3}$ moves are defined for $0<\left\vert
i\right\vert <n$ , and given below:%
\[%
\begin{tabular}
[c]{|c|}\hline
$b_{i}b_{i+1}b_{i}b_{-(i+1)}b_{-i}b_{-(i+1)}\overset{\lambda
}{\longleftrightarrow}1^{6}$\\\hline
\\\hline
$b_{i}b_{i+1}b_{i}b_{-(i+1)}b_{-i}\overset{\lambda}{\longleftrightarrow
}b_{i+1}1^{4}$\\\hline
\\\hline
$b_{i}b_{i+1}b_{i}b_{-(i+1)}\overset{\lambda}{\longleftrightarrow}b_{i+1}%
b_{i}1^{2}$\\\hline
\\\hline
$b_{i}b_{i+1}b_{i}\overset{\lambda}{\longleftrightarrow}b_{i+1}b_{i}b_{i+1}%
$\\\hline
\\\hline
$b_{i}b_{i+1}1^{2}\overset{\lambda}{\longleftrightarrow}b_{i+1}b_{i}%
b_{i+1}b_{-i}$\\\hline
\\\hline
$b_{i}1^{4}\overset{\lambda}{\longleftrightarrow}b_{i+1}b_{i}b_{i+1}%
b_{-i}b_{-(i+1)}$\\\hline
\\\hline
$1^{6}\overset{\lambda}{\longleftrightarrow}b_{i+1}b_{i}b_{i+1}b_{-i}%
b_{-(i+1)}b_{-i}$\\\hline
\end{tabular}
\
\]

\end{definition}

\bigskip

\begin{example}
Two examples of Reidemeister $R_{3}$ are given below:%
\[%
{\includegraphics[
natheight=3.000000in,
natwidth=3.000000in,
height=0.3269in,
width=0.3269in
]%
{rtm23.jpg}%
}
{\includegraphics[
natheight=3.000000in,
natwidth=3.000000in,
height=0.3269in,
width=0.3269in
]%
{rtm13.jpg}%
}
{\includegraphics[
natheight=3.000000in,
natwidth=3.000000in,
height=0.3269in,
width=0.3269in
]%
{rtm23.jpg}%
}
\overset{\lambda}{\longleftrightarrow}%
{\includegraphics[
natheight=3.000000in,
natwidth=3.000000in,
height=0.3269in,
width=0.3269in
]%
{rtm13.jpg}%
}
{\includegraphics[
natheight=3.000000in,
natwidth=3.000000in,
height=0.3269in,
width=0.3269in
]%
{rtm23.jpg}%
}
{\includegraphics[
natheight=3.000000in,
natwidth=3.000000in,
height=0.3269in,
width=0.3269in
]%
{rtm13.jpg}%
}
\]
\bigskip%
\[%
{\includegraphics[
natheight=3.000000in,
natwidth=3.000000in,
height=0.3269in,
width=0.3269in
]%
{rtm13.jpg}%
}
{\includegraphics[
natheight=3.000000in,
natwidth=3.000000in,
height=0.3269in,
width=0.3269in
]%
{rtm23.jpg}%
}
{\includegraphics[
natheight=3.000000in,
natwidth=3.000000in,
height=0.3269in,
width=0.3269in
]%
{rt03.jpg}%
}
{\includegraphics[
natheight=3.000000in,
natwidth=3.000000in,
height=0.3269in,
width=0.3269in
]%
{rt03.jpg}%
}
\overset{\lambda}{\longleftrightarrow}%
{\includegraphics[
natheight=3.000000in,
natwidth=3.000000in,
height=0.3269in,
width=0.3269in
]%
{rtm23.jpg}%
}
{\includegraphics[
natheight=3.000000in,
natwidth=3.000000in,
height=0.3269in,
width=0.3269in
]%
{rtm13.jpg}%
}
{\includegraphics[
natheight=3.000000in,
natwidth=3.000000in,
height=0.3269in,
width=0.3269in
]%
{rtm23.jpg}%
}
{\includegraphics[
natheight=3.000000in,
natwidth=3.000000in,
height=0.3269in,
width=0.3269in
]%
{rtp13.jpg}%
}
\]

\end{example}

\bigskip

\begin{remark}
The number of Reidemeister 3 moves $R_{3}$ is given by%
\[
\text{\# }R_{3}\text{ Moves }=\left\{
\begin{array}
[c]{ccc}%
n\left(  n-2\right)  \left(  6\ell-21\right)  & \text{if} & \ell\geq6\\
&  & \\
n\left(  n-2\right)  \left(  5\ell-16\right)  & \text{if} & \ell=5\\
&  & \\
n\left(  n-2\right)  \left(  3\ell-8\right)  & \text{if} & \ell=4\\
&  & \\
n\left(  n-2\right)  \left(  \ell-2\right)  & \text{if} & \ell=3\\
&  & \\
0 & \text{if} & \ell<3
\end{array}
\right.
\]

\end{remark}

\bigskip

\subsection{Stage 1. (Cont.) The ambient group $A\left(  n,\ell\right)  $ and
the braid mosaic system $\mathcal{B}_{n,\ast}$}

\qquad\bigskip

At this point, we can define what is meant by the ambient group and the
resulting braid mosaic system.

\bigskip

We begin reminding the reader of a fact noted earlier in this paper, namely
the fact that each braid move is a permutation on the set $\mathbb{B}%
^{(n,\ell)}$ of $(n,\ell)$-braid mosaics. \ Thus, since planar isotopy and
Reidemeister moves are permutations, we can make the following definition:

\bigskip

\begin{definition}
We define the ( $(n,\ell)$\textbf{-braid mosaic}) \textbf{ambient group}
$A(n,\ell)$ as the group of all permutations on the set $\mathbb{B}^{(n,\ell
)}$ of $(n,\ell)$-braid mosaics generated by $(n,\ell)$-braid planar isotopy
and Reidemeister moves.
\end{definition}

\bigskip

We need one more definition, before we can move to the objective of this section.

\bigskip

\begin{definition}
We define the \textbf{braid mosaic injection}%
\[
\iota:\mathbb{B}^{(n,\ell)}\longrightarrow\mathbb{B}^{(n,\ell+1)}%
\]
as the map
\[
\beta\longmapsto\beta1
\]
for each $(n,\ell)$-braid mosaic in $\mathbb{B}^{(n,\ell)}$. \ It immediately
follows that the braid mosaic injection induces a monomorphism
\[
\iota:A(n,\ell)\longrightarrow A(n,\ell+1)\text{ }%
\]
from the $(n,\ell)$-braid ambient group $A(n,\ell)$ to the $(n,\ell+1)$-braid
ambient group $A(n,\ell+1)$. \ This monomorphism is called the \textbf{braid
mosaic monomorphism}. \ 
\end{definition}

\bigskip

\begin{definition}
We define an \textbf{braid system} $\mathcal{B}_{n,\ell}=\mathcal{B}\left(
\mathbb{B}^{(n,\ell)},A(n,\ell)\right)  $ of\textbf{ order} $\left(
n,\ell\right)  $ as the pair $\left(  \mathbb{B}^{(n,\ell)},A(n,\ell)\right)
$, where $\mathbb{B}^{(n,\ell)}$ is called the \textbf{set of }$\left(
n,\ell\right)  $-\textbf{braid mosaics}, and where $\mathbb{A}(n,\ell)$ is
called the \textbf{ambient group}. \ Finally, we define a\textbf{ nested motif
system} $\mathcal{B}_{n,\ast}=\mathcal{B}\left(  \mathbb{B}^{(n,\ast
)},A(n,\ast)\right)  $ as the following sequence of sets, groups, injections,
and monomorphisms:%
\[
\mathcal{B}\left(  \mathbb{B}^{(n,1)},A(n,1)\right)  \overset{\iota
}{\longrightarrow}\mathcal{B}\left(  \mathbb{B}^{(n,2)},A(n,2)\right)
\overset{\iota}{\longrightarrow}\cdots\overset{\iota}{\longrightarrow
}\mathcal{B}\left(  \mathbb{B}^{(n,2)},A(n,2)\right)  \overset{\iota
}{\longrightarrow}\cdots
\]

\end{definition}

\bigskip

\subsection{Stage 2. \ Braid mosaic type and braid mosaic invariants}

\qquad\bigskip

Our next objective is to define what it means for two braid mosaics to
represent the same topological braid.

\bigskip

Two braid mosaics $\beta_{1}$ and $\beta_{2}$ of the set $\mathbb{B}%
^{(n,\ell)}$ are said to be of the \textbf{same }$n$\textbf{-braid mosaic
type}, written%
\[
\beta_{1}\underset{n}{\sim}\beta_{2}\text{ ,}%
\]
if there exists an element $g$ of the ambient group $A(n,\ell)$ which takes
$\beta_{1}$ to $\beta_{2}$, i.e., such that%
\[
g\beta_{1}=\beta_{2}\text{ .}%
\]
The braid mosaics $\beta_{1}$ and $\beta_{2}$ are said to be of the
\textbf{same braid mosaic type}, written%
\[
\beta_{1}\sim\beta_{2}\text{ ,}%
\]
if there exists a non-negative integer $k$ such that%
\[
\iota^{k}\beta_{1}\underset{n+k}{\sim}\iota^{k}\beta_{2}\text{ .}%
\]

\bigskip

We now wish to answer the question:

\bigskip

\noindent\textbf{Question}. \ \textit{What is meant by a braid mosaic
invariant?}

\bigskip

\begin{definition}
Let $\mathcal{B}_{n,\ell}=\mathcal{B}\left(  \mathbb{B}^{(n,\ell
)},A(n)\right)  $ be a braid system, and let $\mathbb{D}$ be some yet to be
chosen mathematical domain. \ By an $n$-\textbf{braid mosaic invariant}
$I^{(n)}$, we mean a map
\[
I^{(n)}:\mathbb{B}^{(n,\ell)}\longrightarrow\mathbb{D}%
\]
such that, when two braid mosaics $\beta_{1}$ and $\beta_{2}$ are of the same
$n$-type, i.e., when
\[
\beta_{1}\underset{n}{\sim}\beta_{2}\text{ , }%
\]
then their respective invariants must be equal, i.e.,
\[
I^{(n)}\left(  \beta_{1}\right)  =I^{(n)}\left(  \beta_{2}\right)  \text{ .}%
\]
In other words, $I^{(n)}:\mathbb{B}^{(n,\ell)}\longrightarrow\mathbb{D}$ is a
map that is invariant under the action of the ambient group $A(n)$, i.e.,
\[
I^{(n)}\left(  \beta\right)  =I^{(n)}\left(  g\beta\right)
\]
for all elements of $g$ in $A(n)$.
\end{definition}

\bigskip

\subsection{Stage 3. \ Construction of the corresponding quantum braid system}

\qquad\bigskip

We now use the nested \ braid mosaic system $\mathcal{B}_{n,\ast}$ to
construct a nested sequence of quantum braid mosaic systems $\mathcal{Q}%
_{n,\ast}$.

\bigskip

For pair of non-negative integers $n$ and $\ell$ the corresponding $\left(
n,\ell\right)  $\textbf{-th order} \textbf{quantum braid system}%
\[
\mathcal{Q}_{n,\ell}=\mathcal{Q}\left(  \mathcal{B}^{(n,\ell)},\mathcal{A}%
(n,\ell)\right)
\]
consists of a Hilbert space $\mathcal{B}^{(n,\ell)}$, called the
\textbf{quantum mosaic space}, and a group $\mathcal{A}(n,\ell)$, also called
the \textbf{ambient group}.\ The quantum motif space $\mathcal{B}^{(n,\ell)}$
and the ambient group $\mathcal{A}(n,\ell)$ are defined as follows:\bigskip

\begin{itemize}
\item The\textbf{ quantum motif space} $\mathcal{B}^{(n,\ell)}$ is the Hilbert
space with orthonormal basis%
\[
\left\{  \ \left\vert \beta\right\rangle :\beta\in\mathbb{B}^{(n,\ell
)}\ \right\}  \text{ .}%
\]
The elements of $\mathcal{B}^{(n,\ell)}$ are called \textbf{quantum braids}%
$.$\bigskip

\item The \textbf{ambient group} $\mathcal{A}(n,\ell)$ is the unitary group
acting on the Hilbert space $\mathcal{B}^{(n,\ell)}$ consisting of all linear
transformations of the form%
\[
\left\{  \ \widetilde{g}:\mathcal{B}^{(n,\ell)}\longrightarrow\mathcal{B}%
^{(n,\ell)}:g\in A(n,\ell)\ \right\}  \text{ ,}%
\]
where $\widetilde{g}$ is the linear transformation defined by
\[%
\begin{array}
[c]{rcc}%
\widetilde{g}:\mathcal{B}^{(n,\ell)} & \longrightarrow & \mathcal{B}%
^{(n,\ell)}\\
\left\vert \beta\right\rangle \quad & \longmapsto & \left\vert g\beta
\right\rangle
\end{array}
\]
Since each element $g$ in $A(n,\ell)$ is a permutation, each $\widetilde{g}$
permutes the orthonormal basis $\left\{  \ \left\vert \beta\right\rangle
:\beta\in\mathbb{B}^{(n,\ell)}\ \right\}  $ of $\mathcal{B}^{(n,\ell)}$.
\ Hence, $\widetilde{g}$ is automatically a unitary transformation. \ It
follows that $A(n,\ell)$ and $\mathcal{A}(n,\ell)$ are isomorphic as groups.
\ We will often abuse notation by denoting $\widetilde{g}$ by $g$, and
$\mathcal{A}(n,\ell)$ by $A(n,\ell)$.
\end{itemize}

\bigskip

Next, for each pair of non-negative integers $n$ and $\ell$, let
\[
\iota:\mathcal{B}^{(n,\ell)}\longrightarrow\mathcal{B}^{(n+1,\ell)}%
\]
and%
\[
\iota:\mathcal{A}(n,\ell)\longrightarrow\mathcal{A}(n+1,\ell)
\]
respectively denote the Hilbert space monomorphism and the group monomorhism
induced by the injection%
\[
\iota:\mathbb{B}^{(n,\ell)}\longrightarrow\mathbb{B}^{(n+1,\ell)}%
\]
and the group monomorphism%
\[
\iota:A(n,\ell)\longrightarrow A(n+1,\ell)\text{ .}%
\]

\bigskip

Finally, we define the\textbf{ nested quantum braid system} $\mathcal{Q}%
_{n,\ast}=\mathcal{Q}_{n,\ast}\left(  \mathcal{B}^{(n,\ast)},\mathcal{A}%
(n,\ast)\right)  $ as the following sequence of Hilbert spaces, groups,
Hilbert space monomorphisms, and group monomorphisms:%
\[
\mathcal{Q}_{1,\ell}\left(  \mathcal{B}^{(1,\ell)},\mathcal{A}(1,\ell)\right)
\overset{\iota}{\longrightarrow}\mathcal{Q}_{2,\ell}\left(  \mathcal{B}%
^{(2,\ell)},\mathcal{A}(2,\ell)\right)  \overset{\iota}{\longrightarrow}%
\cdots\overset{\iota}{\longrightarrow}\mathcal{Q}_{n,\ell}\left(
\mathcal{B}^{(n,\ell)},\mathcal{A}(n,\ell)\right)  \overset{\iota
}{\longrightarrow}\cdots
\]

\bigskip

\subsection{Stage 4. Quantum braid equivalence}

\bigskip

Let $\mathcal{Q}_{n,\ell}=\mathcal{Q}\left(  \mathcal{B}^{(n,\ell
)},\mathcal{A}(n,\ell)\right)  $ be a quantum motif system of order $\left(
n,\ell\right)  $.

\bigskip

Two quantum braids $\left\vert \psi_{1}\right\rangle $ and $\left\vert
\psi_{2}\right\rangle $ of the Hilbert space $\mathcal{B}^{(n,\ell)}$ are said
to be of the \textbf{same }$\left(  n,\ell\right)  $\textbf{-braid type},
written%
\[
\left\vert \psi_{1}\right\rangle \underset{n}{\sim}\left\vert \psi
_{2}\right\rangle \text{ ,}%
\]
if there exists an element $g$ of the ambient group $\mathcal{A}(n,\ell)$
which takes $\left\vert \psi_{1}\right\rangle $ to $\left\vert \psi
_{2}\right\rangle $, i.e., such that%
\[
g\left\vert \psi_{1}\right\rangle =\left\vert \psi_{2}\right\rangle \text{ .}%
\]
The quantum motifs $\left\vert \psi_{1}\right\rangle $ and $\left\vert
\psi_{2}\right\rangle $ are said to be of the \textbf{same braid type},
written%
\[
\left\vert \psi_{1}\right\rangle \sim\left\vert \psi_{2}\right\rangle \text{
,}%
\]
if there exists a non-negative integer $m$ such that%
\[
\iota^{m}\left\vert \psi_{1}\right\rangle \underset{n+m}{\sim}\iota
^{m}\left\vert \psi_{2}\right\rangle \text{ .}%
\]

\bigskip

\bigskip

\subsection{Stage 5. Quantum braid invariants as quantum observables}

\qquad\bigskip

We consider the following question:

\bigskip

\noindent\textbf{Question:} \ \textit{What do we mean by a physically
observable quantum braid invariant?}

\bigskip

We answer this question with a definition.

\bigskip

\begin{definition}
Let $\mathcal{Q}_{n,\ell}=\mathcal{Q}\left(  \mathcal{B}^{(n,\ell
)},\mathcal{A}(n,\ell)\right)  $ be a quantum braid system of order $\left(
n,\ell\right)  $, and let $\Omega$ be an observable, i.e., a Hermitian
operator on the Hilbert space $\mathcal{B}^{(n,\ell)}$ of quantum braids.
\ Then $\Omega$ is a \textbf{quantum braid }$\left(  n,\ell\right)
$\textbf{-invariant} provided $\Omega$ is left invariant under the big adjoint
action of the ambient group $\mathcal{A}(n,\ell)$, i.e., provided
\[
U\Omega U^{-1}=\Omega
\]
for all $U$ in $\mathcal{A}(n,\ell)$.
\end{definition}

\bigskip

\begin{proposition}
If%
\[
I^{(n)}:\mathbb{B}^{(n,\ell)}\longrightarrow\mathbb{R}%
\]
is a real valued $\left(  n,\ell\right)  $-braid invariant, then%
\[
\Omega=%
{\displaystyle\sum\limits_{\beta\in\mathbb{B}^{(n,\ell)}}}
I^{(n,\ell)}\left(  \beta\right)  \left\vert \beta\right\rangle \left\langle
\beta\right\vert
\]
is a quantum motif observable which is a quantum motif $\left(  n,\ell\right)
$-invariant.
\end{proposition}

\bigskip

\section{Conclusion}

\bigskip

Much more can be said about this topic. \ \ For more examples of the
application of the quantization procedure discussed in this paper, we refer
the reader to \cite{Lomonaco1, Lomonaco2, Kauffman1, Farhi1}. \ For knot
theory and the braid group, we refer the reader to \cite{Crowell1, Murasugi1,
Kauffman2, Lickorish1, Birman1, Kassel1}; for topological quantum computation,
\cite{Kauffman3, Kauffman4, Kauffman5, Kitaev1,Nayak1, Sarma1}; and for
quantum computation and information, \cite{Nielsen1, Lomonaco4, Lomonaco5}.

\end{document}